\documentstyle[12pt]{article}    
\begin{document}           
\baselineskip=0.33333in
\begin{quote} \raggedleft TAUP 2783-2004
\end{quote}
\vglue 0.5in
\begin{center}{\bf Further Difficulties with the \\
Klein-Gordon Equation}
\end{center} 
\begin{center}E. Comay 
\end{center}
 
\begin{center}
School of Physics and Astronomy \\
Raymond and Beverly Sackler Faculty of Exact Sciences \\
Tel Aviv University \\
Tel Aviv 69978 \\
Israel
\end{center}

Email: elic@tauphy.tau.ac.il
\vglue 0.5in
\vglue 0.5in
\noindent
\vglue 0.2in
\noindent
Abstract:  

The Dirac equation is compared with the Klein-Gordon one. Unlike the
Dirac case, it is proved that the Klein-Gordon equation has problems 
with the Hamiltonian
operator of the Schroedinger picture. A special discussion
of the Pauli-Weisskopf article and that of Feshbach-Villars proves that
their theories of a charged Klein-Gordon particle lack a self consistent
expression for this Hamiltonian. Related difficulties are pointed out.

\newpage

The Klein-Gordon (KG) equation and the Dirac one were published in the
very early days of quantum mechanics (see [1], bottom of pp. 25, 34).
The KG equation is regarded as the relativistic quantum mechanical
equation of a spin-0 massive particle and the Dirac equation describes
a spin-1/2 massive particle. The Dirac equation of an electrically
charged particle can be found in any
textbook on relativistic quantum mechanics and on quantum electrodynamics.
This equation is regarded as a correct description of a system
belonging to the domain of validity[2] of relativistic quantum mechanics.
Thus, for example, the Dirac equation can be used for the hydrogen
atom if one is ready to ignore small effects like the Lamb shift.

Unlike the Dirac equation, the KG equation is not free of objections. In
particular, Dirac maintained his negative opinion on this equation
throughout his life[3]. On the other hand,
claims stating that Dirac's opinion on the KG
equation is wrong were published (see [1], second column of p. 24).

New difficulties with the KG equation were published recently[4]. Thus,
new arguments proving that the KG wave function cannot describe
probability are given; it is proved that a KG particle cannot
interact with electromagnetic fields; the classical limit of the
Yukawa interaction is inconsistent with special relativity and some
other claims.

The 4-current of a particle represents specific properties of its
state, namely its density and its 3-current. 
The KG electromagnetic interaction discussed in [4] relies
on the requirement stating that the 4-current of
a KG particle (like that of any other particle) should not depend
on field variables of {\em external particles}. A further discussion
of this issue is presented near the end
of this work. The present work provides
new arguments that do not rely on this requirement. 
This work examines the structure of the Hamiltonian of the system
in relativistic quantum mechanics. The
significance of the corresponding Lagrangian density is pointed out.

Units where $\hbar = c = 1$ are used. The Lorentz metric
$g_{\mu \nu}$ is diagonal and its entries are 
(1,-1,-1,-1). Greek indices run from 0 to 3 and Latin ones run
from 1 to 3. The summation convention holds for a pair of upper 
and lower indices. The
lower case symbol $_{,\mu}$ denotes the partial differentiation with
respect to $x^\mu$. An upper dot denotes the partial differentiation with
respect to the time. Thus, $\dot \phi \equiv \phi_{,0}$.

Let us examine the theoretical structure of a
Dirac field interacting with an electromagnetic
field. This subject is useful not only for its own sake but also for
the corresponding analysis of the KG equation which is carried out later.
The matter part of the Lagrangian density is (see [5], p. 84)
\begin{equation}
{\mathcal L} = \bar \psi[\gamma ^\mu (i\partial _\mu - eA_\mu) - m]\psi ,
\label{eq:DIRACLD}
\end{equation}
where $\gamma ^\mu $ denotes a set of four Dirac $\gamma $ matrices,
$\psi $ is the Dirac wave
function, $\bar {\psi } = \psi ^\dagger \gamma ^0$
and $\psi ^\dagger$ is the Hermitian conjugate
of $\psi $. The definition
$\gamma ^0 = \beta,\;\; \gamma ^i = \beta \alpha ^i $ relates the
Dirac $\gamma $ matrices and the $\alpha,\; \beta $ ones. The
components of the 4-potential are the electric potential $V$ and the
vector potential ${\bf A}$. Thus, $A^\mu = (V,\bf {A})$.

A variation of  $(\!\!~\ref{eq:DIRACLD})$ with respect to $\bar {\psi}$
yields the Dirac equation (see [5], p. 84) 
\begin{equation}
\gamma ^\mu (i\partial _\mu - eA_\mu)\psi = m\psi .
\label{eq:DIRACEQ}
\end{equation}

An important quantity is the 4-current of the Dirac particle 
\begin{equation}
j^\mu = \bar \psi\gamma ^\mu \psi ,
\label{eq:DIRACJMU}
\end{equation}
which satisfies the conservation law 
\begin{equation}
j^\mu _{,\mu } = 0.
\label{eq:DIRACCONSERVED}
\end{equation}
The validity of this relation is independent of the
external electromagnetic field (see [6], p. 119).
The 0-component of the 4-current $(\!\!~\ref{eq:DIRACJMU})$ represents
the density of the Dirac particle
\begin{equation}
\rho =  \bar \psi\gamma ^0 \psi = \psi ^\dagger \psi.
\label{eq:DIRACRHO}
\end{equation}

The matter part of the
Hamiltonian density is derived from the Lagrangian density
$(\!\!~\ref{eq:DIRACLD})$ by the well known relation (see [5], p. 87)
\begin{eqnarray}
{\mathcal H}  & = & \sum \dot \psi \frac {\partial {\mathcal L}}
{\partial \dot \psi} - {\mathcal L} \nonumber \\
        & = & \psi ^\dagger [\mbox {\boldmath $\alpha $}
\cdot (-i {\mathbf \nabla } -e {\mathbf A}) +
\beta m + eV]\psi ,
\label{eq:DIRACHD}
\end{eqnarray}
where the summation runs on $\dot {\psi }$ and $\dot {\bar \psi }$. (As a
matter of fact, only $\dot {\psi }$ is found in $(\!\!~\ref{eq:DIRACLD})$).
Here quantities should be written in terms of coordinates and 
conjugate momenta. However, this point is not essential for the discussion
carried out below. Hence, it
is skipped throughout this work.

Using the expression for the density $(\!\!~\ref{eq:DIRACRHO})$,
one readily extracts from the Hamiltonian density $(\!\!~\ref{eq:DIRACHD})$
an expression for the Hamiltonian operator used in the Schroedinger's
picture of relativistic quantum mechanics
\begin{equation}
H = \mbox {\boldmath $\alpha $}
\cdot (-i {\mathbf \nabla } -e {\mathbf A}) +
\beta m + eV
\label{eq:DIRACHOP}
\end{equation}

It is well known that the Hamiltonian operator $H$ plays a cardinal
role in the Schroedinger picture of
quantum mechanics, because it defines the time evolution and the
energy states of the system (see [5], p. 6)
\begin{equation}
H\psi = i\frac {\partial \psi}{\partial t}.
\label{eq:HPSI}
\end{equation}
Now, due to the principle of superposition, quantum mechanics uses
equations that are linear in $\psi $. For this reason, the Hamiltonian
operator of $(\!\!~\ref{eq:HPSI})$ should not depend on $\psi $. This
requirement is satisfied by the Dirac Hamiltonian $(\!\!~\ref{eq:DIRACHOP})$.

Substituting the Hamiltonian operator $(\!\!~\ref{eq:DIRACHOP})$ 
into the quantum mechanical relation $(\!\!~\ref{eq:HPSI})$,
one obtains the Hamiltonian form of the Dirac equation
\begin{equation}
[\mbox {\boldmath $\alpha $}
\cdot (-i {\mathbf \nabla } -e {\mathbf A}) +
\beta m + eV]\psi =
i\frac {\partial \psi}{\partial t}. 
\label{eq:DIRACHAMILTONIAN}
\end{equation}
The complete agreement between $(\!\!~\ref{eq:DIRACHAMILTONIAN})$ and
the Dirac equation $(\!\!~\ref{eq:DIRACEQ})$,
derived as the Euler-Lagrange equation
of the Lagrangian density $(\!\!~\ref{eq:DIRACLD})$,
indicates the self-consistence of the theory.

   It is interesting to note relativistic properties of the Hamiltonian
density $(\!\!~\ref{eq:DIRACHD})$ and of the Hamiltonian operator
$(\!\!~\ref{eq:DIRACHOP})$. Examining the first line of 
$(\!\!~\ref{eq:DIRACHD})$ and remembering that the Lagrangian density
${\mathcal L}$ is a Lorentz scalar, one realizes that $(\!\!~\ref{eq:DIRACHD})$
is a tensorial component $T^{00}$ of the second rank tensor
\begin{equation}
T^{\mu \nu } =  \sum  \psi ^{,\mu } \frac {\partial {\mathcal L}}
{\partial \psi _{,\nu }} - {\mathcal L}g^{\mu \nu}.
\label{eq:TMUNU}
\end{equation}
This is the required covariance property of energy density. In classical
physics, energy density is the $T^{00}$ component of the energy-momentum
tensor $T^{\mu \nu}$ (see [7], p. 77). Now, since the probability
density $\rho $ of $(\!\!~\ref{eq:DIRACRHO})$ is a 0-component of a
4-vector, one concludes that also the Hamiltonian operator $H$ of
$(\!\!~\ref{eq:DIRACHOP})$ is a 0-component of a 4-vector. Evidently,
this property is essential for satisfying covariance of
the fundamental quantum mechanical relation 
$(\!\!~\ref{eq:HPSI})$. This discussion shows
just one reason for the usefulness of
constructing the theory on the basis of a Lagrangian density. This
point is used below in the analysis of the Feshbach-Villars 
(FV) Hamiltonian.

It can be concluded that the following properties hold for the Dirac theory:
\begin{itemize}
\item[{1.}] The conserved 4-current depends on $\psi $ and on the
corresponding $\bar \psi $ and is independent of the external
field $A_\mu$.
\item[{2.}] Since the Dirac Lagrangian density $(\!\!~\ref{eq:DIRACLD})$ is
{\em linear} in the time-derivative 
$\partial \psi /\partial t$, the corresponding Hamiltonian
density $(\!\!~\ref{eq:DIRACHD})$ does not contain derivatives of $\psi $
with respect to the time. Hence, in the case of a Dirac particle,
the fundamental quantum mechanical relation 
$(\!\!~\ref{eq:HPSI})$ takes the standard form of an explicit
first order partial
differential equation. Here a derivative with respect to the
time is equated to an expression which is free of time derivatives.
This property does not hold for Hamiltonians that depend on
time derivative operators.
\item[{3.}] The Dirac Hamiltonian operator
$(\!\!~\ref{eq:DIRACHOP})$ is free of
$\psi,\; \bar {\psi}$ and their derivatives. An examination of
$(\!\!~\ref{eq:HPSI})$ proves that this property is consistent 
with the linearity of quantum mechanics and with the superposition
principle as well.  
\item[{4.}] The equation $(\!\!~\ref{eq:DIRACHAMILTONIAN})$ obtained from
the substitution of the Dirac Hamiltonian 
operator $(\!\!~\ref{eq:DIRACHOP})$
into the quantum mechanical relation $(\!\!~\ref{eq:HPSI})$, agrees with
the Dirac equation $(\!\!~\ref{eq:DIRACEQ})$ obtained as the
Euler-Lagrange equation of the 
Lagrangian density $(\!\!~\ref{eq:DIRACLD})$.
\end{itemize}

These four points indicate the self consistency of the Dirac theory.
It is proved below that difficulties arise if one carries out an
analogous analysis of the KG equation.

Let us turn to the Pauli-Weisskopf (PW) 
theory of a charged KG particle (see Section 3 of
[8]). These authors use the Lagrangian density (see eq. (37) therein)
\begin{equation}
{\mathcal L} = (\dot \phi ^* -ieV\phi ^*)(\dot \phi + ieV\phi) -
\sum _{k=1}^3 (\phi _{,k}^* +ieA_k \phi ^*)(\phi _{,k} -ieA_k \phi )
- m^2\phi ^* \phi.
\label{eq:PWLD}
\end{equation}
Note that minor changes are made in the form of
quoted equations. Thus, units where $\hbar =c=1$ are 
introduced; $\phi $ denotes the KG wave function and the 
electromagnetic 4-potential is $A^\mu = (V,{\bf A})$. On the other
hand, the Lorentz metric of quoted formulas is that of the
original articles.

The Hamiltonian density associated with $(\!\!~\ref{eq:PWLD})$ is
written next to this equation (see eq. (37a) therein)
\begin{equation}
{\mathcal H} = (\dot \phi ^* -ieV\phi ^*)(\dot \phi +ieV\phi) +
\sum _{k=1}^3 (\phi _{,k}^* +ieA_k \phi ^*)(\phi _{,k} -ieA_k \phi )
+ m^2\phi ^* \phi.
\label{eq:PWHD}
\end{equation}

The Lagrangian density $(\!\!~\ref{eq:PWLD})$ is used in a derivation
of the second order
equation of motion of a charged KG particle (see eq. (39) therein)
\begin{equation}
(\frac {\partial }{\partial t} -ieV)(\frac {\partial }{\partial t} - ieV)\phi =
\sum _{k=1}^3 (\frac {\partial }{\partial x^k} +ieA_k )
(\frac {\partial }{\partial x^k} +ieA_k )\phi
+ m^2 \phi.
\label{eq:PWEQ}
\end{equation}

The conserved 4-current of this particle is derived too. The 0-component of
this quantity is (see eq. (42) therein)
\begin{equation}
\rho = i(\phi ^* \dot \phi - \dot {\phi }^* \phi) - 2eV\phi ^* \phi.
\label{eq:PWRHO}
\end{equation}
Unlike the case of a Dirac particle, here the 4-current of a KG particle
depends on derivatives of $\phi $ and on external electromagnetic quantities.

Before proceeding with the analysis, let us write down the canonical
Hamiltonian obtained from the application of the first line of 
$(\!\!~\ref{eq:DIRACHD})$ to the Lagrangian density
$(\!\!~\ref{eq:PWLD})$
\begin{equation}
{\mathcal H} = \dot \phi ^* \dot \phi -e^2V^2\phi ^* \phi +
\sum _{k=1}^3 (\phi _{,k}^* +ieA_k \phi ^*)(\phi _{,k} -ieA_k \phi )
+ m^2\phi ^* \phi.
\label{eq:PWHDCAN}
\end{equation}
As mentioned (see [9], p. 68) this expression is not gauge invariant.

Let us examine the issue of the Hamiltonian operator required for
the Schroedinger picture of
the fundamental quantum mechanical relation $(\!\!~\ref{eq:HPSI})$.
It is shown above how easily this task is
accomplished for the Dirac Hamiltonian. In this case one just removes
the Dirac density factor $\psi ^\dagger \psi$ from the Hamiltonian density
$(\!\!~\ref{eq:DIRACHD})$ and extracts the required expression. This
quantity is not given in [8].

The following argument proves that this task can be accomplished
neither for the Hamiltonian density $(\!\!~\ref{eq:PWHD})$ nor
for that of $(\!\!~\ref{eq:PWHDCAN})$. Let $\hat H$ denote the
required operator. Now, apart from multiplicative
factors, the highest order time derivative of 
$(\!\!~\ref{eq:PWHD})$ and $(\!\!~\ref{eq:PWHDCAN})$
 is $\dot \phi ^* \dot \phi $ and that of the density
$(\!\!~\ref{eq:PWRHO})$ is $(\phi ^* \dot \phi - \dot \phi ^* \phi)$.
Hence, $\hat H$ cannot contain the operator
$\partial /\partial t$, because both
$\dot \phi ^* \dot \phi $ and $(\phi ^* \dot \phi - \dot \phi ^* \phi)$
have $\dot \phi $ as the highest order time
derivative of $\phi $. Evidently, due
to the superposition principle and the linearity of quantum
mechanics, $\hat H$ should depend neither on $\phi $, $\phi ^*$ nor
on their derivatives. Hence, under these restrictions on the 
structure of $\hat H$, it is evident that $\hat H$ cannot exist
because $\dot \phi ^* \dot \phi $ is symmetric with respect to $\phi $ and
$\phi ^*$, whereas $(\phi ^* \dot \phi - \dot \phi ^* \phi)$ is 
antisymmetric with respect to these functions. This proof does not
rely on terms containing the electric charge $e$. Hence, it applies
also to the case of an uncharged KG particle described by a complex field.

Let us turn to the theory described in the FV article[1].
These authors construct an expression for the Hamiltonian operator
of a charged KG particle that, in the Schroedinger's picture,
takes the standard quantum mechanical
form $(\!\!~\ref{eq:HPSI})$. For this purpose they use a 2-component
wave function
\begin{equation}
\Psi = \left(
\begin{array}{c}
\psi \\
\chi
\end{array}
\right).
\label{eq:FVPSI}
\end{equation}
where $\psi $ and $\chi $ are linear combinations of the KG wave function
$\phi $ and of $\dot \phi $ (see eqn. (2.11)-(2.17) therein).
These authors present the following Hamiltonian operator that
can be used in the Schroedinger picture of
$(\!\!~\ref{eq:HPSI})$ (see (2.18) therein) 
\begin{equation}
H = (\tau _3 + i\tau _2) (1/2m)({\bf p} -e{\bf A})^2 + m\tau _3 + eV,
\label{eq:FVH}
\end{equation}
where $\tau _2$  and $\tau _3$ are Pauli spin matrices.

The analysis of FV does not rely on a Lagrangian density. Hence,
it is not clear whether or not the Hamiltonian $(\!\!~\ref{eq:FVH})$  
satisfies relativistic covariance. As a matter of fact, a 
proof of this essential property
is not found in [1]. The following analysis explains why it is
impossible to construct such a proof.

Let us analyze covariance properties of $(\!\!~\ref{eq:FVH})$. As stated
earlier, the fundamental quantum mechanical relation $(\!\!~\ref{eq:HPSI})$
indicates that $(\!\!~\ref{eq:FVH})$ should be a 0-component of a
4-vector. The last term of $(\!\!~\ref{eq:FVH})$ is $eV$ where $e$ 
is a Lorentz scalar denoting the charge of the KG particle and $V$ is
the 0-component of the electromagnetic 4-potential $A_\mu $. Hence, the last
term of $(\!\!~\ref{eq:FVH})$ is a 0-component of a 4-vector,
as required. The
second term of $(\!\!~\ref{eq:FVH})$ is $m\tau _3$. Here $m$
is a scalar denoting the KG particle's self mass. Now, the $\tau _3$
Pauli matrix certainly can't transform like a 0-component of a 4-vector.
Therefore, the second term and the last term of $(\!\!~\ref{eq:FVH})$
have different covariant properties.

Furthermore, the
first term of $(\!\!~\ref{eq:FVH})$ is also inconsistent with the last
one, because it is not a 0-component of a 4-vector. 
Indeed, the following expression shows the tensorial form of
this term
\begin{equation}
({\bf p} -e{\bf A})^2 = (E -eV)^2 - (P^\mu - eA^\mu )(P_\mu - eA_\mu )g^{00}. 
\label{eq:FV1}
\end{equation}
Here the first term on the right hand side
is a product of two energy quantities. Hence,
under a Lorentz transformation, $(\!\!~\ref{eq:FV1})$
behaves like a tensorial component $W^{00}$. Here, in
principle, one may alter the tensorial rank of each term by using
the relativistic metric $g_{\mu \nu }$ and the
completely antisymmetric unit tensor of the fourth rank 
$\varepsilon ^{\alpha \beta \gamma \delta }$. Evidently, the rank
of each of these tensors is an even number. Thus, one cannot put the first
term of $(\!\!~\ref{eq:FVH})$, which 
is a component of an even rank tensor $W^{00}$ and the
last one, which belongs to an odd rank tensor, $A^\mu $, 
in the same equation,
without violating covariance. Obviously, the factor $(\tau _3 + i\tau _2 )$
and the Lorentz scalar $1/2m$
cannot settle this contradiction. This
discussion proves that $(\!\!~\ref{eq:FVH})$ violates covariance and
therefore it takes an unacceptable form of the Hamiltonian.

The results of this work are described in the following lines.
First, the theory derived from the Lagrangian density of a charged Dirac
particle is discussed. It is shown that the
Hamiltonian density and the Hamiltonian operator of the
Schroedinger picture are derived in a straightforward manner and
the results are selfconsistent. In particular, the Euler-Lagrange
equation derived from the Lagrangian density agrees with the
fundamental quantum mechanical equation 
$i\partial \psi /\partial t = H\psi $.

It is shown that an analogous structure does not exist for the KG
equation. An expression for the Hamiltonian operator of the Schroedinger
picture is not given in [8] and it is proved above that this quantity
cannot be extracted from the Hamiltonian density
$(\!\!~\ref{eq:PWHD})$. Note also that an
attempt to construct a Hamiltonian operator for a charged KG
particle without relying on a Lagrangian density[1] fails too. 
In this case it is
proved that the suggested Hamiltonian violates
relativistic covariance and should be rejected.

Since no acceptable Hamiltonian operator exists for a charged KG particle,
one obviously cannot close the logical cycle and prove that the Hamiltonian
equation of motion $i\partial \psi/\partial t = H\psi $ 
is consistent with the Euler-Lagrange equation obtained
from the KG Lagrangian density $(\!\!~\ref{eq:PWLD})$. This is
certainly not an easy task, because the KG equation has a second order 
derivative with respect to the time whereas the Hamiltonian density
and the fundamental quantum mechanical equation 
$i\partial \psi /\partial t = H\psi $ contain only first order
derivatives.

The following discussion compares the structure of the Lagrangian
density of the Dirac equation with that of the KG one and 
provides a possible explanation of the
origin of the difficulties of the latter. The unit system where
$\hbar = c = 1$ facilitates this task. Here dimensions of every physical
quantity is written in terms of one unit, which is taken here to be
that of length $[L]$. Thus, energy and momentum have the dimension
of $[L^{-1}]$.

The action $S$ is dimensionless. Thus, the relation
\begin{equation}
dS = (\int {\mathcal L} d^3x)dt
\label{eq:ACTION}
\end{equation}
proves that the dimension of the Lagrangian density is $[L^{-4}]$. The 
first line of $(\!\!~\ref{eq:DIRACHD})$ proves that this is also the
dimension of the Hamiltonian density ${\mathcal H}$. Now, in the case
of the Dirac equation, terms take the first power of
energy, momentum and mass (henceforth called energy-like quantities). 
 By contrast, it is shown
above that this relation holds for the Dirac equation where
$(\!\!~\ref{eq:DIRACEQ})$ is equivalent to 
$(\!\!~\ref{eq:DIRACHAMILTONIAN})$.Hence, the dimension of the product
$\bar \psi \psi $ is $[L^{-3}]$. Thus, in the case of the Dirac
equation terms representing energy-like quantities play a general
role and take the same form for all states of the Dirac particle. On
the other hand $\psi $ and $\bar \psi $ represent {\em specific }
information concerning the particle's state. This is the
underlying reason for the straightforward extraction of the Dirac
Hamiltonian operator $(\!\!~\ref{eq:DIRACHOP})$.

The structure of the KG Lagrangian density $(\!\!~\ref{eq:PWLD})$
(and that of the associated Hamiltonians $(\!\!~\ref{eq:PWHD})$
and $(\!\!~\ref{eq:PWHDCAN})$) differs from that of the Dirac case.
Here energy terms take the second power. Hence, the dimension of
the product $\phi ^* \phi $ is $[L^{-2}]$. Now, since the dimension
of density is $[L^{-3}]$, one finds that in the case of a complex field
of an uncharged particle, the expression for the density is
$i(\phi ^* \dot \phi - \dot \phi ^* \phi )$. A complication arises
in the case of a charged particle $(\!\!~\ref{eq:PWRHO})$,
where the density of the KG particle depends on electromagnetic
quantities too.

Now, in the KG equation of motion $(\!\!~\ref{eq:PWEQ})$, energy-like
quantities do not represent {\em specific}
properties of the KG field but have a {\em general} meaning and take
the same form for all states of the KG particle. On the other hand,
the expression for the density $(\!\!~\ref{eq:PWRHO})$ describes
a {\em specific} property of the field. Thus, energy-momentum
operators $(i\partial /\partial t,-i\nabla)$ play two different
roles in the structure of the KG theory: in the KG equation of
motion they represent energy-momentum and contain no specific
property of the field whereas in the expression for the density
$(\!\!~\ref{eq:PWRHO})$ - which is a {\em specific} property of the
solution - one energy operator changes its role and is used for this
purpose.

The situation becomes even more unexpected in the case where electric
charge and electromagnetic fields are a part of the system. Here
the substitution $P^\mu \rightarrow P^\mu - eA^\mu $ (see [5],
p. 84 and [8], eq. (36)) is performed. Thus, $eA^\mu $, which
is a companion of energy-momentum, is carried together with the latter
and plays a part in the description of the density of the KG particle.

This discussion explains how the KG theory uses energy-momentum
operators for two distinct roles: as energy-momentum operators
representing energy balance in the KG equation, and as a
part of the expression representing a specific property of the
solution, namely its 4-current in general and its density in
particular. This ambiguity is probable the underlying reason for
the inability to extract the KG Hamiltonian operator from the
Hamiltonian density $(\!\!~\ref{eq:PWHD})$.

The following difficulties of the KG equation are discussed above:
\begin{itemize}
\item[{1.}] The theory lacks the Hamiltonian operator required for the
Schroedinger picture.
\item[{2.}]Assuming that this Hamiltonian is constructed, it is not
clear that the second order KG equation is equivalent to the
fundamental quantum mechanical equation $i\partial \psi /\partial t = H\psi $.
\item[{3.}] No justification is given to the dependence of the 4-current
of the KG particle on the external electromagnetic 4-potential.
\item[{4.}] No justification is given to the different meaning of
energy-momentum operators: as energy-momentum operators in the KG
equation and as an element in the description of the 4-current, which
is a property of a specific solution.
\end{itemize}


\newpage
References:
\begin{itemize}
\item[{[1]}] H. Feshbach and F. Villars, Rev. Mod. Phys.
{\bf 30}, 24 (1958).
\item[{[2]}] F. Rohrlich {\em Classical Charged Particles}
(Addison-wesley, Reading Mass, 1965). pp. 3-6. 
\item[{[3]}] P. A. M. Dirac {\em Mathematical Foundations of Quantum
Theory}, Editor A. R. Marlow (Academic, New York, 1978). (See pp. 3,4). 
\item[{[4]}] E. Comay, Apeiron, {\bf 11}, No. 3, 1 (2004).
\item[{[5]}] J. D. Bjorken and S. D. Drell {\em Relativistic Quantum
Fields} (McGraw, New York, 1965).(See chapter 12.)
\item[{[6]}] V. B. Berestetskii, E. M. Lifshitz and L. P. Pitaevskii,
{\em Quantum Electrodynamics} (Pergamon, Oxford, 1982).
\item[{[7]}] L. D. Landau and E. M. Lifshitz, {\em The Classical
Theory of Fields} (Pergamon, Oxford, 1975).
\item[{[8]}] W. Pauli and V. Weisskopf, Helv. Phys. Acta, {\bf 7}, 709 (1934).
English translation: A. I. Miller {\em Early Quantum
Electrodynamics} (University Press, Cambridge, 1994). pp. 188-205.
\item[{[9]}] G. Wentzel, {\em Quantum Theory of Fields} (Interscience,
New York, 1949).

\end{itemize}

\end{document}